**The Impact Of World War I On Relativity:  Part III, The Aftermath**

**ABSTRACT**

     Neither the world nor science came to an end when the gunfire stopped on 11 November 1918 (close to 11 AM in some time zone), but neither would ever be the same again.  Part I of this inquiry (Observatory 138, 46-58, April 2018) looked at the development of general relativity under the Rubric of Gerald Holton's "Only Einstein; only there; only then."  Part II (Observatory 138, 98-116, June, 2018) addressed the activities, relativistic, classical, and otherwise of many (mostly) physicists who were interacting with Einstein, working on relativistic gravity, or, sometimes, against it, and leaving tracks that can still be followed.  Part III considers some of what happened to Einstein, his theory of gravity, and related science after the war and, perhaps, because of it.  A subset of the items will probably be familiar – the 1919 eclipse expedition and the founding of the International Astronomical Union the same year; Einstein's 1921 Nobel Prize (for the discovery of the law of the photoelectric effect).  Others perhaps less so, including a flood of books about GR (pro and con) with the end of paper rationing surely playing a role; AE's 1922 trip to Paris, and the gory details, swings and roundabouts of gravitational radiation/waves and the cosmological constant.  It is left as an exercise for the reader to decide which items are primarily scientific and which primarily political.  The long-range issues of "is general relativity the right theory of gravity?" and "do we have better wars?" come at the end.  And I am going to start in a slightly improbable place.

Virginia Trimble
Department of Physics and Astronomy
University of California
Irvine, California 92697-4575, USA
    and
Queen Jadwiga Observatory
Rzepiennik Biskup, Poland

vtrimble@uci.edu
phone: 949-824-6948
fax: 949-824-2174

**INTRODUCTION**

In the summer of 1921, a 26 year old, newly minted MD traveled by train from Moscow to Berlin, getting hung up briefly at the Lithuanian border. In Berlin, he conceived the idea of a peace-promoting project of publishing, in both the original languages and in Hebrew, two volumes of recent significant papers by European Jewish authors, one eventually devoted to Orientalia and Judaica, the other to Mathematics and Physics. This second volume of the *Scripta Universitates Atque Bibliotecae Hierosolymitarum* was partially edited by Albert Einstein; included the Einstein and Grommer 1922 paper[1]; and, as the rapidly-aging young man later explained, had been rather difficult to assemble, because many French savants did not care to be involved in a project in which there would also be German participants. There was, in fact, only one French chapter, by Hadamard (of the transform). Others came from Tulio Levi-CIvita, Theodor von Karman, H. Bohr (not Niels, but his brother, a mathematician), S. Brodetzky (uncle of the late Leon Mestel), a Landau (not Lev) at Göttingen, a hyphenated Popper (not Daniel Magnes) at Vienna, a somewhat mysterious Loewy of Frankfurt (later metamorphed into Cornelius Lanczos of Dublin), and several others whose names I did not recognize.

The young man's father paid for the publications, out of rapidly-declining resources, and they thereby played a role in the establishment of the Hebrew University in Jerusalem, Einstein's visit to which you met in Part I, because the volumes could be exchanged (in what was then a common custom) for volumes published by other universities, giving the library a start.

The polymathic MD, who later practiced as a psychiatrist, emigrated to the Palestinian Mandate in 1933 and to the US in 1939. From 1946 to 1955, he again interacted sporadically with Einstein in Princeton. Near the end of this period, he gave AE the first half of what would become his best-known and most contentious publication. Some of the more objectionable passages, to which AE took exception, were thereby removed before the volume in question saw light of print, though it was still sufficiently contrary to the known laws of physics to engage a distinguished Harvard astronomer in violent opposition, and to force a change of publishers to MacMillan, which had few technical books on its books and so could afford to annoy the scientific community. The Harvard pundit required a younger female colleague to provide a review of the book which was also very negative.

If you haven't yet guessed that the pundit was Harlow Shapley and the younger colleague Cecilia Payne (Gaposchkin), please go to Ref. (2) to identify the Einstein-mentored author. It was the "Venus" section that Einstein had seen. I read the author's later volumes, *Ages in Chaos* and *Oedipus and Ahkenaten* when they were new, but you are probably too young even to have heard of them.

Surprisingly at least to me, in his last, April 1955 interview with I. Bernard Cohen, two weeks before his death, Einstein chose to address Velikovsky and *Worlds in Collision* (neither by name). He said that both book and person were "crazy" but not "bad," and regretted that the American scientific community had tried to prevent publication of the book. The figure is the same one that appeared in Part I, with focus now shifted to the outcomes.

I have not found on rational order in which to present the pieces of the aftermath and so have grouped them under cutesy-poo section headings.

**FORTUNE, FILMS, AND FLOOD ON FOLIOS**

Actually the fortunes involved were very modest. As the war ended, the shortage of money and food mentioned in a number of the letters[3] did not immediately end. The Allies maintained their blockade and were slow in fulfilling a promise to prevent starvation (Doc. 664, 665, and notes thereto, early December 1918). Einstein of course won the 1921 Nobel Prize in Physics (for "discovering the law of the photoelectric effect") given in 1922, but the money went to his divorced first wife as he had promised as far back as June 1918 (Doc. 562). Perhaps worth noting are that she would have control only over the interest, not the capital; that in case of her death or remarriage, the full sum would go to their sons; and that AE expected the Prize to be more than 40,000 German marks.

Luckily the prize was in Swedish krona, since the German mark went through dire inflation in the early 1920s, saved by Hjalmar Horace Greeley Schacht. You have to love the name, whatever you think of the person. His parents had been in the United States when Horace Greeley (of "Go West, young man," and he meant Pittsburgh) was the democratic candidate defeated by Ulysses S. Grant in 1872. Schacht also survived WWII.

Just how much was the Prize worth? In 1920, each was 134,100 Swedish crowns, down 10% or so from the pre-War value, the equivalent of US $36,250 or £8,252[4]. Circumstances have increased the recent prizes to of order a million US $. On the other hand, in 1915, a gallon of milk cost $0.36[5]. You could hire an unskilled laborer for $1 per day (Trimble family lore) and a skilled astronomer for $1 per hour.

Mileva Maric Einstein died in 1948 (after AE's second wife, Elsa in1936). Elder son Hans Albert became a successful engineer, fairly distant from his father, whom he outlived, as did younger son Eduard (d. 1955), who, however, spent much of his adult life in mental health asylums.

Motion pictures intended to educate are not new (nor, it has to be said, typically very successful). In 1922, Hanns Walter Kornblum (1878 – 1970) produced a 2 or 3 hour German film explaining (mostly) special relativity, with bending of light at the end, though it was originally intended to cover all of special and general relativity. It had a large cartoon component and does not survive, though a 30 minute English language version may[6]. A 1923 American cartoon, produced by Max Fleischer and intended to explain GR can be found on YouTube, in my case by chance. Fleischer was also the producer of the Betty Boop cartoons, including "Betty Boop. Cinderella. Two-color" which takes less than 10 minutes to view, has better tunes than other Cinderella films, and is an excellent illustration of how two rather than three primary colors can produce attractive effects.

Some combination of enhanced fame after the 1919 eclipse results (later section), the challenges of understanding what Einstein had done, and perhaps also a general quest for royalties in the wake of the war and subsequent economic turmoil, unleashed an enormous flurry of books about/for/against/explaining GR. Freundlich[7] led off in 1916. Einstein got into the act the next year[8]. In English we got Whitehead[9], Eddington twice over[10,11], Birkhoff[12], and Hermann Weyl in translation.[13] Ludwig Silberstein, who had tackled the special theory in 1914[14] came back a decade later on the general theory.[15]

There were more, naturally, in German. Goenner[16] and Gutfreund and Renn[17] have assembled a sizable list, surely not exhaustive. Here are just the years and authors:

| 1917 | (Moritz Schlick) |
|------|------------------|
| 1918 | (Wilhelm Wien, Werner Bloch) |
| 1919 | (Moritz Schlick, Jan Arnoldus Shouten, in Dutch) |
| 1920 | (Hans Thirring, Max Born, Alexander Pflüger, Harry Schmidt, Max Hesse, Hans Reichenbach, Ernst Cassirer) |
| 1921 | (Felix Auerbach, Alexander Moszkowski, Max von Laue two volumes, Hans Thirring, August Kopff, Wolfgang Pauli – encyclopedia chapter |
| 1922 | (Paul Gruner, Max Born, Ernst Richard Neumann |
| 1923 | (Karl Vogtherr, Lorentz et al.[18]) |

There were also contemporaneous volumes harshly critical of relativity and of Einstein himself by Hugo Dingler (1921), Philippe Lenard (1920, 1921), and Johannes Stark (1922). It seems likely that Sten Lithigius, writing in Swedish[19a] would have had particular impact on the Nobel physics committee, but Friedman[4], devotes his whole chapter 4 to "Einstein must never get a Nobel Prize." And one cannot read Ref. 17 or any other modern discussion of the history of GR without hearing repeatedly of the role of anti-Semitism in the German (and some other) reactions to relativity, and, for that matter, to quantum mechanics.

On the positive side, by 1922-24, Alexander Friedmann, Cornelius Lanczos, Enrico Fermi, and Eli Cartan were creeping into the journal literature with papers important enough to be cited by Misner, Thorne, & Wheeler.[19] Lest I forget to mention it elsewhere, Kasner[20] wrote in this same time frame to explain why you could not put a bunch of 4 (space-time) dimensional universes side by side in 5-dimensional space. In six, apparently you can.

Einstein's fame has never waned. Time Magazine declared him the person of the Century (meaning the 20[th]), on December 31[st], 1999, primarily for relativity, though other topics were mentioned. He still inspires strange sorts of enthusiasm, being featured in the February, 2018 issues of *1843* (an adjunct to *The Economist*) as a "sartorial role model who combines substance and style" in a piece called "best dressed".[20a] The "stylish" items were a brown leather jacket recently bought at auction for $149,700 and the absence of socks. And science historical Helge Kragh[20b] has created an imaginary, November 1928 oral history interview of Einstein by Kragh's imaginary uncle Carl Christian Nielsen (imaginarily 1887 – 1971) dealing primarily with cosmology. The chapter is accompanied by an apparently real photograph of AE on a park-like bench with Arthur Eddington, who was imaginarily interviewed by Nielsen on 2 December 1938.

Another Ph.D. physicist named Frank Potter has put forward another set of imagined interviews with physicists of the past, available only on Kindle. Of the fifty, Einstein gets four, Galileo and Feynman only three each.

But the enthusiasm for general relativity waned. In the fall of 1919, Charles G. Abbot (Home Secretary of the US National Academy of Sciences) told George Ellery Hale that everybody would be heartily sick of relativity by April, 1920. Indeed the Hale lecture that year was the Curtis-Shapley debate on "The Distance Scale of the Universe," though Abbot had proposed the causes of the ice ages or some topic in zoology or biology. The short life of the IAU Committee on Relativity follows shortly. And the flood of GR books slowed to a trickle, only monographs by Otto Heckmann and Peter Bergmann appearing in the 1940s.[20c,20d]

Einstein's own enthusiasms apparently also somewhat waned! W.W. Campbell, director of Lick Observatory, led a 1922 expedition to a solar eclipse in Australia[23]. He wrote in due course to Einstein, reporting their results, (considerably more definitive than the 1919 numbers). A response came, which is preserved in the Lick archives,[*] expressing Prof. Einstein's "cordial gratitude and transmitting his admiration for the extraordinary diligence and accurateness of measurements taken." But it is signed "The Secretary," though Einstein had been writing enormous numbers of his own letters just a few years before.

**IMMEDIATE SEQUELS: THE 1919 ECLIPSE AND THE IAU**

Arthur Stanley Eddington (1882-1944) was a Quaker and pacifist, who had several near-misses with trouble during the Great War, while he had been Secretary of the Royal Astronomical Society.[21,22] In that capacity, he received letters and papers from Wilhelm de Sitter (who appears in Part II and below, in connection with the cosmological constant). He was the only one of those you will meet in this section young enough to be at risk of conscription. Eddington was initially deferred because of the importance of his work, but called up in 1918, and he asked for conscientious objector status (legal, but not regarded as honorable by most of his contemporaries). Intervention by Frank Watson Dyson, the Astronomer Royal, and others, kept him out of prison. Dyson and the others here all have entries in ref. 23.

Eddington and Dyson both recognized that an eclipse was coming on 29 May 1918, when the sun would be projected against the star-rich Hyades cluster. They were the primary organizers. The Royal Greenwich Observatory expedition (observers Charles R. Davidson, 1875 - 1970, and Andrew Crommelin, 1865 - 1939, born in Northern Ireland) went to Sobral, Brazil. The Cambridge expedition under Eddington went to Principe Island off the coast of Africa. Dyson's RGO had lost 36 members of staff to active duty during the war, and work fell behind, though he hired retirees, refugees from Belgium, conscientious objectors, and WOMEN in their places.

Getting the plates home, measuring them, and deciding what the star positions meant all took time. There have been sporadic fusses about whether the published data were completely honest, but the announcement of results equal to the prediction of general relativity by Dyson, Eddington, and Davidson[24] led to headlines splashed across the New York Times and elsewhere, and made Albert Einstein a superstar. The accuracy of their result does not matter to our present understanding of gravity, for the observations have been repeated many times, optically at many other eclipses (down to the 21 August 2017 one in the US[25]). Radio astronomy took over when it was noticed that strong compact sources 3C 273 and 3C 279 would pass behind the sun each October, and the bending of both light and radio waves is as close to the GR value as technology can make it[26].

The founding of the International Astronomical Union, whose centenary is fast approaching, was also a direct outcome of the great war. For its story let us turn to Adriaan Blaauw[27] (1914 – 2010), one of the founders of the European Southern Observatory, the first chair of the Board of Directors of *Astronomy and Astrophysics*, and the president of the IAU, (1976 – 1979) who shepherded the return of the People's Republic of China to membership without loss of the astronomers from the Republic of China (Taiwan) with a rubric, "one nation, two adhering organizations," adopted afterwards by others of the ICSU Unions.

---

[*] Ms. Ilse Ungeheuer, of the current Lick staff, sent me a copy, and I confess to having found it surprisingly unenthusiastic.

George Ellery Hale[28] was the founder of three observatories, each in its day with the world's largest telescope, Yerkes, Mt. Wilson, and Palomar Mountain. He was also co-founding editor of the *Astrophysical Journal*, and in 1903 – 1904 he wrote to a number of "men of science" interested in solar research[29], inquiring whether they thought some sort of international organization on the topic would be useful. Acting upon their positive responses, he arranged to be chair of a committee of the US National Academy of Sciences, and, in that capacity, wrote to 17 national academies and scientific societies inviting them to send representatives in 1904 to the St. Louis Exposition. Sixteen sent representatives, Prussia refusing, but some Germans came from the German Physical Society.

They agreed to meet again in Oxford in September 1905 and to establish an International Union for Cooperation in Solar Research. This Solar Union was approved in 1907 by the International Association of Academies (held up perhaps by the Prussians). The IAA last met in 1913 in St. Petersburgh. The Solar Union again convened in 1907 in Paris (Meudon), in 1910 in Pasadena (and at Mt. Wilson, where participants were duly impressed by the 60" telescope), and where they agreed to expand their remit to include stellar research, especially astrophysics (meaning, in those days, spectroscopy).

The last fully international astronomical meeting before the First World War was still called the Solar Union for short, and happened in Bonn July 30 to August 5, 1913. Then there was a war. Well before it ended, indeed before the United States entered, the NAS offered to organize the scientific resources of the country in preparation for war. Woodrow Wilson ("He kept us out of war" having gotten him re-elected that same year) accepted the offer, and the National Research Council came into being in April with Hale as chairman (as well as Foreign Secretary of the Academy). It included representatives of educational and research organizations, industrial and engineering research, technical bureaus of the Army and Navy, and government representatives. Hale regarded the NRC as a sort of model for an international organization to be established after the war ended. He never seems to have much doubted the outcome.

Following a great deal of toing and froing (mostly letters and telegrams, but some sea voyages)[27], there took place the First Inter-Allied Conference on the Future of International Organizations in Science, at Burlington House, London 9-12 October 1918, followed by a second conference in Paris November 26–29. Participants came from Belgium, Brazil, France, Great Britain, Italy, France, Portugal, the United States, and Serbia.

Several points in the resolutions adopted at these conferences echo down to the present. First that the nations at war with the Central Powers withdraw from the existing conventions relating to international Scientific Associations…as soon as circumstances permit. Second, that the new associations be established without delay by the nations at war with the Central Powers with the eventual cooperation of neutral nations. Third that certain associations, such as the Metric Convention, be taken into consideration during the peace negotiations (a sample of these follows shortly).

At the Paris meeting, the name International Research Council was accepted, and it acquired a council with Picard (France) as president, Schuster (BR) as general secretary, and Hale (US), Lecointe (BE), and Volterra (IT) as vice-presidents. Astronomy was clearly well represented. The first formal assembly of the IRC took place in Brussels 18 – 28 July 1919.  Represented were Belgium (about half the participating scientists), France, the US, Great Britain, Canada, New Zealand, Poland, Rumania, and Serbia, again with many astronomers.  Additional countries from the Allied side immediately entitled to join the IRC and Unions under it were Australia, Brazil, South Africa, Greece, Japan, and Portugal.  The founding IAU members were Belgium, Canada, France, Greece, Italy, Japan, United Kingdom, and the United States[30].  The neutral countries invited at Brussels to join the IRC and then the various unions were China, Siam, Czecho-Slovakia, Argentine Republic, Chile, Denmark, Spain, Mexico, Monaco, Norway, Holland, Sweden and Switzerland.

The early additions to the IAU were Mexico (1921), Czechoslovakia, Denmark, Norway, Poland, Rumania, Spain, and the Netherlands (all 1922 at the 2nd General Assembly), Switzerland (1923), Portugal (1924), Egypt and Sweden (1925), Argentina (1927), Vatican City (1932), China, USSR, and Yugoslavia (1935), South Africa (1938), and Australia (1939).  And then, as you might just barely recall, there was another war. After 1996, all the former republics of the USSR were deemed to have inherited her right to membership, if they could pay the dues*, while the Yugoslavian right went to Croatia, and Serbia (with Herzegovina) did not adhere until 2003.

Hungary, Germany, Austria, and Turkey (chronologically) all belong to the post world war II period, along with a number of other countries (somewhat fluctuating) as and when they felt the need to support their indigenous astronomers and the ability to pay the dues.  The US share in 1948 was 2500 gold Francs or about $748.  It is more now.

*Treaties, Conventions, and Agreements allowed to survive the Great War.*  These appear in Articles 282-287 of the Versailles Treaty (my copy of which once belonged to a certain Frank M. Mason).  These are about 36, and a few of them are subject to Germany fulfilling certain stipulations.  You must go to the original document to see the complete list, but here are a few of my favorites, some of which have echoes down to the present, and violations of some of which occurred during the lead up to WWII (each has attached dates, 1857 – 1913, most in the 80s and 90s; and places where the agreements were signed, e.g. Vienna, Washington, Rome, St. Petersburg, Lisbon): Protection of submarine cables; sealing of railway trucks subject to customs inspection (Lennin not mentioned); unification of commercial statistics; guaranteeing free use of the Suez Canal; suppression of nightwork for women (oops, there go our astronomers); suppression of white phosphorous in matches; suppression of the White Slave Trade (oops, there go our..); unification and improvement of the metric system (kilogram still to be sorted out in 2018 or later); unification of pharmacopaeial formulae for potent drugs (still an issue!); concert pitch; precautions against phylloxera (save our wine!!); protection of birds useful to agriculture (bees not mentioned); Postal Union and Telegraphic Conventions; fisheries in the North Sea outside territorial waters (again still an issue in many places!).

Of course the new arrangements did not go through unopposed!  Kapteyn objected initially to any exclusion of neutrals, and when they were invited in, he tried to discourage the Dutch Academy from adhering for as long as

---

* You will have to take my word that I am now typing these, in an order determined by geography, not any alphabet, from memory.  Lithuania, Latvia, Estonia; Belarus, Moldava, Ukraine; Armenia, Georgia, Azerbaijan; Kazahkstan, Kirghistan, Tajikistan, Turkmenistan, Uzbekistan, and Umbrellastand, otherwise known as Russia.

Germany was excluded[27]. Be grateful he failed on that one, since Jan Oort was an enormously valuable officer and member for many years! He has by far the largest number of index entries in Blaauw's history.

The most bitter objections came from German astronomers[32], Struve ending his "On the development of German astronomy" with "Per aspera ad astra" . The Astronomische Gesellschaft had been in the habit of thinking of itself as "the" international astronomical society, and with some justification. From its 1863 founding through 1918 60% of the astronomers who passed through as members were from outside Germany, including many from the US, UK, France, Italy, Poland, Russia, and so forth. These included (with years of membership, d indicating that was also the year of death): George Ellery Hale himself (1893 d. 1938), Eddington (1913 d. 1944), W.W. Campbell (1891 d. 1938), F.W. Dyson (1906 d. 1939) E.C. Pickering (1877 d. 1919), Kapteyn (1887 d. 1922), both Curtis (1910 d. 1942) and Shapley (1925 – d. 1945) of the Great Debate, de Sitter (1909 d. 1934), also Georges Lecointe of Belgium (1908 – 1921) and Vito Voltera of Italy (1898 – 1921), founding vice-presidents of the IRC, Eduard Benjamin Baillaud (1877 – 1921) founding president of the IAU, Svante Elis Strömgren, founding head of the IAU Central Bureau for Telegrams (1900 – 1945). If you care to go back further, you will also find John Couch Adams (also Galle who did find Neptune, but he was German) and Simon Newcomb. Karl Schwarzschild (1896 d. 1916), many of the astronomers Hilmar Duerbeck identified as having served for Germany, some killed in WWI, were also AG members, as was Albert Einstein (1921 – 1933), and our old friends Baron Lorand Etovos (1898 d. 1919) and Erwin Fritz Finlay-Freundlich (1913 – 1926).

About 50% of the (smaller number of) members who passed through AG in 1919 – 1945 were still from other countries, Sweden and the US dominating. But this dropped to about 15% after the Second World War and has remained low. If you sum Russia and the USSR, they have contributed the largest number of foreign members, followed by the US and Sweden. An alternative sum of Austria plus Hungary plus Austria-Hungary actually wins with close to 9% of the integrated membership. The female representation started to grow from near zero around 1920 and is now a smidge more than 10%.

In the event, some of the astronomical responsibilities that had resided in Germany before WWI, including portions of the *Carte du Ciel* and the central telegraph bureau, were moved elsewhere. Variable stars, the compiling of minor planet data, and the maintenance of the astronomical bibliography were not relocated until after the Second war[27].

It has sometimes been written, somewhat incorrectly, and probably even by me, that the death of Hale's Solar Union and the establishment of the International Astronomical Union occurred under the Treaty of Versailles. In fact the only astronomical item there (yeah, I read the whole thing) is in article 131, which says:

> Germany undertakes to restore to China within twelve months from the coming into force of the present Treaty all the astronomical instruments which her troops in 1900 – 1901 carried away from China, and to defray all expenses which may be incurred in effecting such restoration, including the expenses of dismounting, packing, transporting, insurance and installation in Peking.

I had very much doubted that this ever occurred, and hadn't quite realized that the removal was part of a much larger looting of Chinese possessions in the wake of the Boxer Rebellion. In fact, Prof. Lu Lingfeng of the University of Science and Technology in China e-informed me that the instruments, probably eight, were returned. They were things like armillary spheres, sextants, quadrants, sun dials, and celestial globes, all large, bronze, mostly supported by dragons (also bronze), and partially dating back to the 1600's when Jesuit astronomers were in China. They are now in the Beijing Ancient Observatory, which has a web presence.

The International Astronomical Union began its life with many traces of Hale's Solar Union, including triennial general assemblies, more than one official language (English and French, German having been dropped from the Solar three), and committees, later commissions to focus on specific territories and tasks. The last new one in the Solar Union had been classification of stellar spectra. The proposal to broaden from the sun to other stars is generally credited to Karl Schwarzschild, but the topic had been on the agenda before the meeting started, and was introduced by Hugh Frank Newall of Cambridge. The formal motion came from Schwarzschild in German, immediately after he claimed his English was not good enough for the purpose.[**]

The IAU also began its life with 32 Committees,[27] each with a president from one of the founding nations. Four were solar-oriented (though Hale was president of only one). And Committee Number 1 was Relativity (ah!! Here we are back on topic) under A.S. Eddington. It voted itself out of existence at the 1925 General Assembly in Rome, and Relativity did not reappear at the IAU until the 1970 General Assembly at Brighton (UK), where Commission 47 (Cosmology) and 48 (High Energy Astrophysics) were blessed and established.

Other Solar Union relics included, in 1919, nations and their academies and societies as the adhering organizations. Individual human beings as members finally appeared in revised by-laws in 1958 (the Solar Union considered this step, but firmly rejected it), and we now outnumber the national adhering organizations 100:1 or thereabouts. And in the latest iteration of Divisions and Commissions, it is not entirely clear where General Relativity belongs.

## SCIENTIFIC ISSUES THAT LINGERED

There are (at least) three of these: the reality of what Einstein wrote as lowercase $\lambda$ and we write aa upper case $\Lambda$, the cosmological constant; whether gravitational waves (radiation) can carry energy. And is GR the right theory of gravity? We think we know the answer to all three: yes, yes and no, but here are some additional steps on the paths from the early days. The relevant chapters from Gutfreund and Renns are "The Genesis of Relativistic Cosmology" 5 and 6 "The controversy over gravitational waves."[32]

Lambda has a history something like the American folk dance, "The Hokey Pokey" (You put your left foot in, you put your left foot out, you put your left foot in and you shake it all about. You do the hokey pokey and you turn yourself around; that's what it's all about." Try singing this with "lambda" instead of "left foot."). If you have already heard some version of the story and are tired of it, feel free to skip to a later section.

Einstein's well-advertised original motivation for introduction of the extra term in his field equations[33] was the desire for a static universe. At various times he also noted, as you have surely been told, that it could be thought of as the second integration constant of a second order differential equation (Hubble's H being the first).  In principle, there are two such static solutions, called spherical (where all geodesics will pass through two poles) and elliptical (where the geodesics intersect only once).  Because one must not think of the latter as looking like a 3-dimensional ellipse (Doc. 300), it is perhaps better not to think of it at all.  The two differ by a factor two in volume for a universe with a given value of density or Λ.  AE explains this most clearly in Doc. 300 to Freundlich, who had drawn his attention to that sort of geometry.  Felix Klein enters the story with Doc. 319[3].  Other participants in the exchanges included de Sitter and Weyl.

Both Einstein's initial cosmology and the empty "De Sitter hyperboloidworld" emerge in extended debate-by-letter among the four (see p. 351-372 and the associated letters in Ref. 3).  De Sitter space did not have the singularity Einstein "accused" it of (merely an artefact of coordinate choice).  But Einstein's static universe really is unstable, and collapses or expands in response to any perturbation. Various sources credit several different contributors for demonstrating this instability.  But I started with a more serious worry – aren't systems generally perturbed from outside? Not to worry.  Tolman (sect. 159 of Ref. 34) shows the basic calculations and then tells his readers that, if free radiation condenses into matter or freely moving particles get captured by condensation, the model will start to expand.  Conversely if matter transforms into radiation (stars do a lot of this) the model would start to contract.  We can, therefore, turn with a clear conscience to Friedmann and Lemaître.

Alexander Alexandrovich Friedmann[35] (1888 – 1925), whose father, also Alexander Alexandrovich Friedmann was a ballet dancer and musician, interrupted masters-level study to serve in WWI in aviation units of the Army on the northern and southern fronts.  Soviet scientists were able to catch up on western European scientific advances only after the end of the war and their revolution, at which point Friedmann set out to study general relativity.[43]  The first Russian survey of the topic came from AAF's friend and colleague V.K. Frederiks (a joint volume[42] appeared only after AAF's death; but there had been a 1923 book *Theory of Relativity* (sorry my typewriter doesn't speak Russian) by Yakov Ilyich Frenkel (father of the middle author of ref. 35).

Can we still connect up with that period?  Yes, if "we" are quite old!  Vladimir A. Fock, who led the Russian delegation when they walked out of the meeting of GR6 in Copenhagen in summer 1971 had been part of a seminar group with which AAF discussed cosmological ideas; and George Gamow (1904 – 1968) had just started work on cosmology with AAF when the latter died, and so Gamow completed a 1928 thesis on what we would now call barrier penetration in alpha decay.  The last chapter of ref. 35 makes clear just how unpopular cosmology was in the Soviet Union until about 1962.  One wonders whether a longer life for Friedmann and Gamow's remaining in Leningrad could have made a difference.  It is usual to blame the decline of cosmology there on Lev Landau (I've done so myself), but Tropp et al.[35] point out that Landau and Lifshitz "gave an exemplary presentation of Friedmann's cosmology in their famous *Course of Theoretical Physics*."

Just what was that cosmology?  Friedmann showed that there are solutions of the Einstein equations for a homogeneous universe, both with and without Λ that can either expand or contract, as different functions a(t) depending on relative values of density of mass-energy and of Λ[36,37].  Does all this contradict whatever you might have previously heard about evolutionary cosmologies violating materialistic principles of Communism?  Never mind. The "antis" put all the blame for an expanding universe on the "reactionary scientists Lemaître, Milne, and others." (p. 223-224 of ref. 35).

So what then of Georges Henri-Joseph-Edouard Lemaître (1894 – 196)? He also interrupted his studies (at the Catholic university in Louvain, Belgium, in engineering) when called to serve as an artillery officer. Post-war, he completed a first degree in mathematics and physics, wandered among Cambridge (UK), Harvard, and MIT, writing a thesis in French that included a form of what we now call the Tolman-Oppenheimer-Volkoff equation of state (useful for neutron stars), and receiving a 1927 PhD from Louvain. Meanwhile, however, he had enrolled at the seminary at Malines, Belgium and was priested in 1923. This was not, present Louvain astronomers tell me, a reaction to the Great War, but something he had always planned.

Lemaître's pioneering paper[38] definitely favored an expanding universe with a non-zero cosmological constant and a very dense state at its origin. He demonstrated the instability of Einstein's static universe, used Slipher's galaxy redshifts to estimate what we now call the Hubble constant at 600 km/sec/Mpc, interpreted $\Lambda$ as a vacuum energy density, described the early universe as a "primeval atom" (meaning the mass of a few billion galaxies all at nuclear density), and suggested that cosmic rays were a remnant of that primordial state[46]. Though we would now disagree with some of the details, one really has to agree that the Abbe was the "father of the Big Bang"[40,41]. Unfortunately the 1927 paper appeared in a Belgian journal not much read in the UK, the US, or Russia, and the version of his paper published in *Monthly Notices*[45] had the expansion constant calculation removed, with his own acquiescence, as being of no "actual" importance, a confusion in meaning between French actuel ("current") and the similar-sounding English word.[46]

In later years, there was some Soviet work, described as deriving from the Friedmann solutions[35]. I mention only a few names of mathematicians and physicists who might be familiar to you in other contexts. Matvei Petrovich Bronshtein (one of many executed in 1937), O.D. Khvolson (who as Chwolson published the very first gravitational lensing paper.)[47a] A.A. Belopolsky (who influenced Gerasimoch and so Ambartsumian indirectly) and, of course, Landau & Lifshitz, who explored both sign conventions – positive $ds^2$ = time-like (my choice) and space-like (Ref. 27).

We bid temporary farewell to Einstein, who had described $\Lambda$ as something to be determined by observations of the distribution of stars and such (Doc. 325 from 1917 in ref. 3) and on another occasion as the second integration constant (Doc. 591). Famously, he backed away from $\Lambda$ when he accepted that the universe expands, somewhere around April 1931.[47]

Erwin Schroedinger (1887 – 1961) pops in here, before turning to his equation and his cat. He had been called up into active service as an artillery officer for three years and then was transferred to meteorology.[48] Often the greatest risk was boredom, and he filled large notebooks with calculations, but also received a citation "for his fearlessness and calmness in the face of recurrent heavy enemy artillery fire." Back on civilian soil, he turned his attention briefly to relativistic universes and came out in favor of the cosmological constant[48] and held by it to the end.[50,44] He outlived Einstein by about six years, and their disagreements (more often about unified theories but also about $\Lambda$) continued throughout their lives.[44]

Was Λ ever without an astronomical supporter? Eddington held the fort until 1944; Schrödinger until 1961; Lemaître until 1966. Soon after that, Gerard Henri de Vaucouleurs (1918 – 1995) maintaining a value of the Hubble constant near 100 km/sec/Mpc required a cosmological constant to make the universe old enough for its contents[52,53] pretty much until his death, when large scale structure folks[54] took over.

You know how the story turns out – with 2011 Nobel Prize in Physics going to Perlmutter, Riess, and Schmidt for discovery of cosmic acceleration (that is, significant non-zero Λ and the current best buy universe having 70% or so of its energy density (positive through the pressure is negative) in Λ or dark energy, or quintessence, or whatever you want to call it. And we can bridge the gap from the last of those who held on beyond Einstein to universe-2018. One of Neta Bahcall's early studies of very large scale distribution of galaxies[54] pointed out that the data were easier to understand with the help of a cosmological constant. A plodding review of all possible DM candidates as understood in 1987[55] included as a dark mimic a cosmological constant which Λ=1 could provide Ω=1 without dark matter. G as a function of length scale was the other mimic. And the third bridge seems to have left no paper trail.

One of the symposia that was part of the IAU General Assembly in Kyoto in 1997 concerned cosmology and ended with a panel discussion on the cosmological parameters. This did not make it into the proceedings but is high on my list of memorable events, because the organizers recognized at the last minute that they had empaneled eight men and so added me. A couple of the panelists, including "Chip" Arp were not subscribers to the conventional hot big bang universe and so declined to choose parameters. But leading off for the conventional view was J.P. Ostriker of Princeton, who said that H was about 75, the universe flat, and about 1/3 of the mass-energy in matter of some sort and 2/3 in cosmological constant. When my turn came, I said I agreed with Jerry, except that my H was a bit smaller (disciple of Sandage!) and my Λ a bit larger. And a majority of the panelists agreed that some cosmological constant was needed to make the universe older than its oldest stars for any likely H and to model most successfully the formation of large scale structure. None of us received Nobel Prizes for this!

**THE REALITY AND PROPERTIES OF GRAVITATIONAL WAVES / RADIATION**

The two words mean the same thing in this context, though "radiation" is perhaps firmer in saying that they carry energy. But it is one of those scary words, like nuclear (especially when pronounced noocooler), and the billion pound gorilla, LIGO, declared that they are gravitational waves, preferably not to be confused with gravity waves, which happen places like the earth's atmosphere and have gravity as the restoring force (in contrast to sound, which has pressure as the restoring force).

Within Newtonian gravitation, information is propagated instantly. If the sun vanishes, we fly off immediately, not after 8 minutes. But as early as 1905, Henri Poincaré[56] pointed out that Lorentz transformation required ("…la propagation des las gravitation n'est pas instantee, mais se fait avec la vitesse de la lumiere"), that gravitation travel, at a finite speed, that of light. Next on the field was Max Abraham (who appears briefly in Part II), whose own theory of gravitation was once regarded by Einstein as a viable alternative to GR, but later repudiated. Abraham wrote[59] that gravity could have no analogue to electromagnetic waves because a gravitational dipole would have the sum of the inertial masses and the acceleration equal to zero. That is, waves might be valid solutions of the field equations, but there would be no way to generate them.

Einstein's first statement on the subject dates also from 1913 (Collected papers vol. 4 No.18, p.229), and was a response to a question from Max Born about how fast the effect of gravitation propagates. At the same speed as light AE said, for infinitesimal disturbances of the metric. The next person to ask was Karl Schwarzschild (whom you also met in Part II), writing from the Russian front to ask about waves in Einstein's theory (he had already correctly calculated the perihelionic precession of Mercury), in a communication that does not survive. Einstein's response (Vol. 8, Doc. 194), was that relativistic gravitation would have no waves analogous to electromagnetic ones. But his first paper on the subject[57] came within the same year.

Lest we once again do the Hokey Pokey, this time sticking our right hands in and out, let me refer you to Chapter 7 of ref 17 for some of the details, though they seem to have missed the denial of reality from Levi Civita[60] in 1917, even before AE's more comprehensive discussion[58]. It is perhaps not a coincidence that he was president of the IAU Committee on Relativity when it voted itself out of existence.

From 1918 to 1937, Einstein was apparently not interested in gravitational waves, or anyhow not interested enough to publish on the subject. Arthur S. Eddington (of the eclipse), stepped up to the spinning cricket bat[61,62], defended the reality of the waves and their ability to carry energy and provided the factor of the two needed to correct AE's quadrupole formula. He did not, however reach a firm conclusion on whether the orbit of a pair of masses would decay owing to the emission of gravitational waves.

The difference between Eddington's spinning rod and his binary star is that the former has forces and energies that are not just due to gravitation. That difference remained key to the reality disputes that continued beyond 1923 and, believe it or not, have still not quite ended.[*]

Einstein pops back into our story in 1937 with the then young Nathan Rosen (1900-1995), in an encounter with the publications process that has since become modestly famous. Kennefick[64] provides the most complete version, but here is a precis. The paper as originally written claimed that there could be no energy-transporting waves in GR. They submitted it to *Physical Review*, in which AE had already published since coming to the United States. The editor (Tate), sent the paper to a reviewer, later revealed as H.P. Robertson 1903-1961, of the Robertson-Walker metric. Robertson found serious errors in the calculations and relayed them to the editor who informed Einstein that the paper could not be accepted in its present form. AE was deeply angered, writing that he had sent the paper to be published, not criticized, and withdrawing it. Back at Princeton, where Robertson was until 1947, he discussed the calculations with HPR, who was able in person to persuade Einstein (and Rosen, who was, however, just then in the Soviet Union), to correct the calculations and revise the paper. But *Physical Review* never saw hide-nor-hair of AE again, and the paper[65] appeared in the Journal of the Franklin Institute of Philadelphia, still in 1937.

---

[*] A sphere of uniform density or density varying only with radius is a monopole. We have lots of approximate mass monopoles in the universe and indeed live on one. The expansion or contraction of a monopole yields no radiation whether the sphere is charged or massive or both. A uniform sphere of magnetic north, or a point, would be a magnetic monopole; we find none of those, and the lowest order EM radiation is dipole, when the distributional of charges changes in some more complex way then expansion or contraction of changes in some more complex way than expansion or contractions of a sphere, for instance a plus and a minus charge dancing the Hokey Pokey. Weber[63] assures us in his Eqn. 7.36 that the lowest-order multipole gravitational radiation is quadrupole. assures us that Eqn. 7.36 that the lowest-order, multipole gravitational radiation is quadrupole. You are supposed to remember that the most functions can be expanded in multipoles, and to save you from having to look it up, here is Eqn. 7.36. $\int T_{il} d^3x = \frac{1}{2} \left[ \int \int T_{00} x^i x^l d^3x \right]_{,00}$ "OK, Another way to say it is that for an isolated oscillating system, the dipole moment vanishes as a consequence of conservation of linear momentum, which is equivalent to what Abraham wrote. And yet another verbal version from Gutfreund & Renn[17] "Gravitational waves are produced in leading order by a mass source changing along two perpendicular directions, for instance a weight-lifter doing squats."

Rosen wrote an additional gravitational wave paper from the Soviet Union and another after he had relocated to Israel (cited by Weber[63]), on some of the technical difficulties with sources and propagation. Later in life he turned to non-GR, bimetric theories of gravitation[66] and was the president of the International Society on General Relativity and Gravitation the year (1974), we met in Israel.

Rosen could possibly hold some record for length of time from first to last paper on a topic, from 1937 to 1993, when he and a young colleague showed carefully that, for a cylindrical gravitational wave in empty space, the energy and momentum densities were positive and "reasonable"[70]. He had noted this back in 1958, promised further details, but was slow in providing them, for reasons, he wrote, that he had long forgotten.

Leopold Infeld (1898-1968), of Einstein, Infeld, Hoffman, carried on with anti-wave (or at any rate anti-energy-transport), papers from the 1930's at least until 1960 as he moved from the US to Canada and back to Poland where he had been born (well, it wasn't Poland then, but you know what I mean). The early papers were single-author, some later ones had student co-authors (including the fairly well known Plebanski, Schild, and Michalska-Trautman)[66,67].

Improbable as it may seem, "wave denialists" have persisted not only past the discovery and analysis of PSR 1913+16 (the Hulse-Taylor[68] binary radiator), but even beyond the LIGO announcements[69]. Each press release from the latter has provoked a "no such thing" response from A. Loinger and T. Marsico of Milan, starting with ref.[70].

But to return to the mainstream* revival of interest in "existence and nature" of gravitational radiation paralleled that the revival of general of relativity in general (ugh, sorry). Significant events were the 1955 Bern conference[71] which had been intended to honor Einstein on the 50th anniversary of his "miraculous year", but ended up mourning him; the Chapel Hill conference[72] in 1957, organized by Bryce and Cecile DeWitt, which counts as GR1; and the 1959 Royaumond Conference[73]. At this last, Peter Bergmann said it would be unfair to vote on the reality of the radiation in the absence of Leopold Infeld (who had been at Bern, and spoke against). He also said it would be a major advance if anything came of the "schemes" of Joseph Weber.

Names connected with gradually –improving calculations, leading to gradually increased confidence that the energy and momentum content of the waves was positive and, as Infeld said, "reasonable", include Hermann Bondi, William Bonner, Felix Pirani, Ivor Robinson, and John A Wheeler and Joseph Weber[74]. Particle physicists attach a good deal of importance to an argument from Richard Feynman which they call "beads sliding on a wire," but this clearly has non-gravitational forces and so does not respond to the difficulties perceived by the late denialists, and, indeed, by Bill Bonnor himself.

---

*Retournes a nous moutons suggests either that we all follow the scientific leaders like sheep or like Handel's sheep, all go astray.

Ok!  Let's see if we can sort out what was being argued about.  The continuing problem was that, although Einstein's equations have wave solutions, a pseudotensor* for energy and momentum was zero (I don't know whether this is the same objection as that of Loinger, that particles all follow geodesics and so cannot be carrying energy in waves).  At the Chapel Hill conference, Infeld [74a] expressed his on-going objections.  In the summary talk, Bergmann wrote that Weber and Wheeler[74] concur that waves don't carry any energy in the case of cylindrical waves.  He wasn't sure whether there would be spherical wave solutions, let alone how you could generate them from oscillating quandrupoles.  Equally unclear was whether an orbiting pair of point masses would lose energy at a rate given by the square of an amplitude.

But this is the wrong way to look at the problem.  Weber & Wheeler note in passing that a closed universe has no total energy and zero curvature, and that electromagnetic radiation would seem non-existent because it wiggles a test particle one way and back again to the same state, so that no energy was absorbed?  No, because the wiggling charge itself emits EM radiation – the radiation or back reaction – and so drains the passing waves.  One should look at gravitational waves the same way.  A test particle is moved by the passing wave, and the invariant space-time interval between two test particles is changed.  They in turn send out gravitational information as a radiation reaction, so energy has been drained from the wave.

This approach leads rather naturally to thinking of test masses as detectors and expressing the result of passing waves as the rate of change in separation to that separation, $\Delta s/s = h$.  The radiation appears only in a third approximation to exact solutions, with "advanced" potentials in the calculation, and the motion of the test particle(s) is transverse to the passing wave.  The proper description, therefore is not "ripples in space time" but "transverse shear strains of the spacetime metric"[75].

My take on how it all played out appears at greater length in Ref. 67.

## IS GENERAL RELATIVITY THE RIGHT THEORY OF GRAVITY?

"No, because it is not a quantum theory and cannot be made into one" is the answer one has heard for many years.  Very crudely, the issue is that, if you try to renormalize GR in the way that Quantum Electrodynamics deals with electric charges and their interactions, you can hoke up finite answers in the first-order corrections ("one loop" approximation), but the others all come out larger, not smaller, so the procedure blows up instead of converging.

Einstein himself expected that, just as GR had supplemented or supplanted Newtonian gravitation and mechanics, GR itself would someday be superseded by a better, more complete theory (Ref. 3, Doc. 323). *  Even at that time, he probably had in mind some unified theory of gravitation and electromagnetism, though his first paper moving in that direction came five years later.  Meanwhile, he at least expressed interest in the upcoming 1919 solar eclipse (Ref. 3, Doc. 486), as an additional GR test.

---

*That bothersome pseudotensor appears somewhere in Landau & Lifshitz; in R.C. Tolman 1930 Phys Rev 35, 875, a paper by Chr. Møller, and elsewhere.

Has such an improved theory turned up so far?  No, or you would have heard about it.  Conversely, you may or may not have read items claiming that there is no necessity, since relativistic and quantum mechanical effects appear in such different contexts (so wrote Freeman J. Dyson a while back in *New York Review of Books*). The very early universe, boiling away of primordial black holes, and near the centers of other black holes would seem to be counterexamples, but I have not visited any of these.

## RECENT SUPPORT AND TESTS

Does gravitation travel at the speed of light?  The first answer to this came from the advance of the perihelion of Mercury.  For which "getting the right answer" says that $v_g = c$ to within 5% or so.  There was a brief flurry of worry that some neutrinos were faster than light[76] which almost as quickly as light went away.  Or perhaps light was faster than gravity[77], which, said the authors, would solve the "horizon" and "causality" problems of standard big bang cosmology with no need for inflation.  If this were right, then the slope of the spectrum of primordial density fluctuations would be 0.96478 (vs. 1.0 for the Harrison-Zeldovich spectrum).  The authors asserted that adopting their proposal would "inform quantum gravity."  But, we can now skip directly to the LIGO binary neutron star event (of 17 August 2017), with gamma rays arriving 1.7 seconds after the gravitational wave burst[78].  This sets the two speeds the same to within $10^{-15}$ and the mass of the graviton at less than $10^{-54}$ grams [79]. We are still far from the Fritz Zwicky limit of $10^{-63}$ gram, which follows if there is no higher-order clustering of galaxies[80].  Confidence that the speed of gravity is close to that of light, or anyhow much larger than the speed of earthquake waves through ground and soil is such that it has been proposed to use the waves radiated by shifts of ground as an early-warning system for quakes[81].

---

*AE wrote, 4 April 1917 to Felix Klein    No matter how we draw a complex from nature for simplicity's sake, its theoretical treatment will ultimately never prove to be (adequately) right.  Newton's theory for ex.  seems to describe the gravitational field completely with the potential $\varphi$.  This description proves to be insufficient, the guv functions take its place.  But I do not doubt that the day will come when his approach will also have to give way to a principally different one for reasons that we do not anticipate today.  I believe that this process of securing the theory has no limits. I am sending you my last paper together with these lines.  The gist of its content is in particular, that the size of the universe seems to be linked to the mean density of matter.  It is not at all out of the question that in the foreseeable future the statistics of fixed stars will confirm or refute the theory.

And to David Hilbert 15 Nov 2015  "…since I often racked my brains to construct a bridge between gravitation and electromagnetism"….I am tired out and plagued with stomach pains besides.

---

How precise is the equivalence principle?  This is another topic to which the LIGO double-neutron-star event has made and will make further limits possible (Ref. [78] and references therein).  Meanwhile, the weak equivalence principle is tested by dropping Galileo…no, wait, dropping massive objects of different mass and composition in a vacuum to see whether they land at the same time (in air they do not, but you can approximate the real experiment either with two pendula of identical length and different bob masses or by dropping a sturdy book, held horizontally, and with a smaller piece of paper on it so the air can't get to it).  The MICROSCOPE experiment[82] used a hollow platinum-alloy cylinder centered inside a hollow titanium-alloy cylinder in space.  First results say that inertial and gravitational masses are equal to a part in $10^{14}$.  The goal, with additional data to be analyzed, is a part in $10^{15}$.

The strong equivalence principle, also held by Einstein to be essential to his theory, says that the part of the mass of an object that is due to its own self-gravitation should also have inertial and gravitational masses equal.  Most terrestrial objects (even your department head), have modest self-gravity, but nature has given us pulsar PSR J0337+1715, with one white dwarf in close orbit with it, and another white dwarf further away.  If the pulsar and its close companion (having different percentages of self-gravitational mass-energy), fell at different speeds toward the distant WD, this would show up as a precession of the orbit, and a periodic change in the pulsar timing.  None has been seen[83] to within about 2 parts in $10^6$.

If it bothers you that the constraint on the strong principle is weaker than the constraint on the weak principle, please pause for a glass of Cinzano Bianco (ice, no lemon, please in mine), and rejoin us for the miserable collection of ideas in the next section.

## ALTERNATIVE THEORIES OF GRAVITATION AND COSMOLOGY

The number of these has been countably infinite, some predating or contemporaneous with GR, with brief appearances in Parts I and II, a sprinkling from the 1920s, 30s, 40s, 50s and so forth, with no end in sight, even if you ignore ideas that start with strings, branes, self-reproducing inflation and other ideas part of modern theoretical physics.  Steady State or its modifications is probably best known[84].  I suppose it will vanish with the last of its founders and supporters, the youngest of whom is slightly older than I.  Others are associated with the names of P.A.M. Dirac, E.A. Milne, Hans Alfven, Irving Segal, Roland Omnes, Oskar Klein, M. Milgrom, Jacob Bekenstein, and others best remembered for other contributions, even the much-lauded Arthur S. Eddington[85].  Many recent alternatives have among their goals the elimination of the need for dark matter.

Keep an eye out (perhaps that third one on the tops of our reptilian heads), for Ref. 86, a chapter for which I was invited to provide, but couldn't manage to reach agreement with the CEO on how many theories to include.  I, of course, wanted very many, at least in a table with dominant properties, rather than extended examination of a few.

So, by way of compensation, you get here only two very recent ones.  First Donald Lynden-Bell (whose passing in February 2018 I mention with deep sorrow), and S.M. Chitre asked in these very pages[87], "Does viscosity turn inflation into the cosmic microwave background and $\Lambda$?"  The answer "yes" yields a total volume for the universe of 55777 $(c/H_0)^3$ or about $2.25 \times 10^{34}$ $pc^3$.

Second, Andre Maeder of the University of Geneva has proposed[88] "A new model, based on the dynamical effects of the scale invariance of the empty space: the fall of dark matter." Dark matter is replaced by a slight effect of scale invariance on Newton's laws; inflation is replaced by the effect on Einstein's equations. And "the scale invariance of the empty space is also present in the fundamental theory of electromagnetism."

The test of a new theory remains, however, the ability to reproduce all the good features of the previous theory while still making new predictions or accounting for old observations that were previously puzzling. From that point of view, the situation has not changed since the years of refs. 89 and 90, when one had to admit that General Relativity has passed all the tests thrown at it, better than various competing theories, including some intended to lead the way to quantum gravity and superunification.

## WHAT BECAME OF ALBERT EINSTEIN?

Well, like the hero of every biography, he dies at the end. But let's look at a few items along the way, beginning with the paper trail as he moves away from the quantum ideas he pioneered and eventually away from the main stream in other ways. Here are my favorite five:

1. The Einstein A and B coefficients,[91] the derivation of the relationship among which was a mainstay of qualifying exams in the days when physicists were supposed to think about atoms. You are too young to remember this, but it was one of the very few items on my first, failed three-hour oral qualifying exam that I got right.
2. His generous, surely unprecedented and rarely followed reading, editing, and submitting of papers by Satyendra Bose, containing what we now call Bose-Einstein statistics.[92]
3. The provocative question, "Can Quantum Mechanical Description of Physical Reality be considered complete?"[93] Their answer was "no," and may well in some deep sense have been the right answer. But quantum mechanics has in common with general relativity that, if you follow the rules and do a calculation, the results always agree with experimental and observational data. Whether this counts as "understanding" is up to you.
4. One of many attempts at understanding motion in general relativity, sometimes mentioned as AE's last "useful" paper.[94]
5. An attempt to use kinetic energy of moving point masses to prevent the sort of collapse that Oppenheimer and Snyder[95] had reported.[96] This feels to me like a sort of flying off the handle upon encountering something one doesn't like. I've done it; perhaps you have too. Not being Einsteins prevents us from having our loose screws appear instantly in high-repute journals. Email and online sites allow us to be foolish even faster.

Moving forward, Einstein's scientific endeavors increasingly focused on attempts to unify gravitational and electromagnetic forces, even after the recognition of a nuclear force. He said[97] that it was his experience with the theory of gravitation that determined his expectations. That is, a long struggle, with moments of despair and rejoicing was to be expected, leading to eventual success. Erwin Schrödinger also spent many of his later years hunting for some theory that would unify the forces,[98] with equal unsuccess.

The number of people working on various forms of unified field theory, or theory of everything, now greatly exceeds two. It is not 100% certain that their collective scientific creativity exceeds that of Einstein + Schrödinger, but they have much more powerful tools of strings, branes, and multiverses at their disposal. It is, however, pretty much guaranteed that any unified field theory that might emerge and triumph will be a quantum one, which would presumably have pleased Erwin but not Albert.

### THE EVENTS OF 1922-23

There have been whole chapters and books written about Einstein's April, 1922 trip to Paris.[99,100] This was the second half of a two-part visit originally arranged for 1914 by Paul Langevin, whose lab had worked on sonar during WWI. The first part came off pleasantly. The 1922 part included a public pairing of talks, variously described as a discussion or debate, between Einstein and Philosopher Henri Bergson (1859-1941)[*]

Walther Ratenau was a strong advocate for Einstein making the trip in hopes of mending relations among European scientists; not all his Berlin colleagues agreed. And Langevin had had to work very hard to make the Paris side of the visit come off.[101]

The speakers genuinely disagreed about the nature of time. Their dialogue is published in the July 1922 issues of *Bulletin de la Societe Francaise de Philosophie*. AE maintained that there were only two sorts of time, psychological (like his remark about 10 minutes spent sitting on a hot stove vs. 10 minutes next to a pretty woman), and the time of physics, hosted in equations. HB maintained that there is also philosophical time, to which AE said, "Il n'ya donc un temps des philosophes." Topper and Canales agree that the two didn't understand each other very well. Jimena Canales is scheduled to speak on 3 October 2018 at the American Center for Physics in College Park, Maryland on "The trouble with Einstein's time" in the Lyne Starling Trimble Lecture Series (yes, my father).

My answer to "what time is it?" is "about half past 2.725 K," and high time I finished Part III. This answer has now been available, with increasing precision, since 1965. I have no idea how Einstein would have reacted to it, but Prof. Canales apparently doesn't find it satisfying, or she would not still be lecturing about the topic.

Einstein and Bergson agreed about the merits of attempting European scientific reconciliation, and served together on a League of Nations international commission on intellectual cooperation (chaired by Bergson, and including Marie Curie[102]). They disagreed about religion and the role of government, Einstein having written to Rathenau (Ref. 3, Doc. 305) that the only proper roles of nation-states were to look after hospitals, universities, the police, and so forth, for which some of the Swiss cantons were too small, but most European nations far too large.

---

[*] Bergson was the son of a Polish-Jewish father and British-Jewish mother. He became president of the British Society for Psychical Research 1913. He wrote in his 1937 will that he thought Catholicism was an appropriate complement to Judaism, but did not convert, because he didn't want to be seen to be escaping the events befalling Jews. The Vichy government offered him exemption from having all his offices and titles taken away from him, but he resigned these rather than accepting.

The Nobel Prize events also belong to 1922-23. Of 32 nominations for 1921, 14 were for A.E. (Friedman Ref. 4 p. 129). Many of the scientists entitled to enter nominations did not. The Swedish Academy voted not to award the 1921 prize. In 1922 they voted for Einstein for 1921 and Bohr for 1922, with the ceremony to take place in December, 1922 in Stockholm.

Einstein was in Japan (he picked up his prize in Gothenberg in 1923, lecturing on relativity, though the prize was for the photoelectric effect). His trip was in response to a request from a Japanese publisher for lectures on relativity in June 1922, and somewhat motivated by death threats he had received after Rathenau's assassination. En route back, the Einsteins stopped in Palestine, where he spoke at the site that was to become the Hebrew University, beginning in Hebrew, continuing in French, and ending in German. Details of the trip appear in the recently published "Travel Diaries",[103] reviewed in *Science* (360, 722, 2018) by Andrew Robinson.

Also newly to hand is the latest Volume 15: The Berlin Years: Writing & Correspondence June 1925-May 1927. I haven't read it yet, but a review[104] mentions how very active Einstein was, interacting with colleagues on scientific and organizational issues. He "applied for grants, refereed papers; administered funds and institutions; grappled with personal issues, and was bored in meetings."

The letters, documents, and all have become so numerous that the paper publication has many items only in a Calendar of Abstracts. I pluck out one item, because it leads us directly to the next and last section. "The 1925 Locarno Treaties renewed Einstein's optimism in the prospects for European reconciliation."

Remember Great War hostilities ended in a June, 1919 Treaty of Versailles (the Allies and Germany, the US signing through never implementing its commitment therein to the League of Nations). Over the next year, similar "agreements" took in Austria-Hungary, Bulgaria, and Turkey, none with the US as a party (though there were subsequent US-Central Powers treaties), and Turkey refusing to sign off on hers.

The 1925 Locarno (Switzerland) Treaties (there were seven) aimed at solidifying the western borders of France and Belgium with Germany (with the Ruhr by then back on the German side), Great Britain and Italy acting as guarantors. The price was leaving the eastern borders with Poland and Czechoslovakia relatively unprotected.

**LONG-TERM IMPACT**

Do we have better science? Certainly we have models, explanations, unexplained data, covering a much wider range of phenomena than did our scientific great grandfathers of 1914-18. It is much less obvious that there is more, or even equal, space for individual geniuses, to the point where the awardees of Nobel, Kavli, Breakthrough, Dan David, Gruber, Ambartsumian and similar prizes have begun to recognize entities like "A, B, C, and the D Team," though the Nobel holds its fortress at three. War, near occasions of war, and fear of war have unquestionably funded and driven many of these expansions. Martin Harwit[105] has worried that vitally significant science may somehow have been missed as a result of this process, though he gave no examples of, for instance, near misses.

The gravest result of WWI and its settlement was, of course, World War II, and some modern historians have suggested that the whole thing should just be described as the 31-year war, part 2 starting at the flimsy boundary left at Locarno. Do we have better wars, perhaps, at least different in the sense of being so far self-limiting, like common colds compared to the Black Death, and restricted in area involved compared to WWII, though 73 years is not very long in the great scheme of things.

As for impact on general relativity, three very important outcomes of WWII were radar giving rise to radio astronomy, German rocketry giving rise to X and gamma-ray astronomy from space, and (counting the lead up, the war and the aftermath) massive relocations of physicists.

Radio astronomy has given us not just better measurements of light deflection by the sun and large numbers of discrete sources that could be counted to rule out steady state but also the cosmic microwave background radiation (absolute time in the universe) and the first quasars. X-ray astronomy gave us binary systems with black hole components, whose behavior has on the whole confirmed the Schwarzschild and Kerr solutions of Einstein's equations. Various combinations of X-ray, gamma-ray, and radio data (plus long-suffering optical astronomy, some using adaptive optics developed for military purposes) have told us that most massive galaxies have black holes at their centers with masses a bit less than $10^{-3}$ of the stellar mass, and that black hole birth and accretion are accompanied by relativistic jets that can point at various angles to the line of sight.

As for the relocation of people, Einstein, Weyl, and Peter Bergmann to Princeton; Bondi and Gold to England; and Schrödinger and Lanczos to Ireland are the golden tip of an iceberg. The founders of the Texas Symposia on Relativistic Astrophysics, Ivor Robinson, Alfred Schild, and Engelburt Schucking were all born places other than Texas, indeed places other than the US. Leopold Infeld was described in one of the web sources I encountered as, in his day, Canada's greatest theoretical physicist. Aspects of the Cold War sent him journeying again, along with Nathan Rosen, David Bohm, and Bernt Peters, a cosmic ray physicist who had worked with Oppenheimer and ended up in Denmark.

Newspapermen used to speak of "the Afghanistan effect," meaning that three million people killed in an earthquake someplace distant and obscure would get fewer column inches than a lost dog in the neighborhood. Growth, indeed overgrowth, of instantaneous communication has reduced this effect, leaving us all far more aware of battles of other places and other peoples. No one quite knows what will be the weapons of World War III. But World War IV will be fought with stones, so said Einstein in 1949. This is already beginning to happen on the border of Israel and Gaza, which he had once hoped might be a homeland for both the peoples who claimed it.

**ACKNOWLEDGEMENTS**

I am once again enormously grateful to Diana Kormos Buchwald of the Einstein Papers Project at Caltech for copies of the 1914-1918 volumes and for advice and counsel along the way. Daniel Kennefick of the University of Arkansas has been the official reviewer. Philip Helbig caught German and other errors. The Tiresome Task of Turning Typed Text to Twenty first-century Technology was undertaken by Ms. Alison Lara and Ms. Jan Strudwick, who are women of valor whose price is beyond that of rubies (and yes, I think Einstein would have recognized the quotation; Diana surely will and she is one, too).